\documentclass[12pt]{article}
\usepackage{fleqn,espcrc1}
\usepackage{epsfig}

\newcommand{\AmS}{{\protect\the\textfont2
  A\kern-.1667em\lower.5ex\hbox{M}\kern-.125emS}}

\begin{document}
\title{Introduction to Effective Field Theory}

\author{Barry R. Holstein\address{Department of Physics-LGRT\\ 
        University of Massachusetts, Amherst, MA 01003}}

\maketitle

\begin{abstract}

\end{abstract}

\section{Introduction}

The basic idea behind effective field theory (EFT) methods is to describe 
nature in terms of a fully consistent quantum field theory, but one 
which is valid in a limited energy range.  Now strictly speaking this means
that almost any theory is an EFT except string theories that try to 
be a theory of everything (TOE).  Indeed even the quintessential quantum 
field theory---quantum electrodynamics---is an EFT in that it must break
down at the very highest energies.  In fact we represent the high energy
contribution in terms of a conterterm which, together with the divergent 
component from the loop integration, we fit in terms of the experimental
mass and charge.

In order to get a better feel for this idea it is useful to consider a
simpler example from the regime of ordinary quantum mechanics---that 
of Rayleigh scattering, or why the sky is blue.  
\subsection{Rayleigh Scattering}

Before proceeding to QCD, let's first examine effective field theory
in the simpler context of ordinary quantum mechanics, in order to get familiar 
with the idea.
Specifically, we examine the question of why the sky is blue, whose answer can be
found in an analysis of the scattering of photons from the sun 
by atoms in the atmosphere---Compton scattering.\cite{skb}  First we examine the 
problem using traditional quantum mechanics and, for simplicity, consider 
elastic (Rayleigh) scattering from
single-electron (hydrogen) atoms.  The appropriate Hamiltonian is then
\begin{equation}
H={(\vec{p}-e\vec{A})^2\over 2m}+e\phi
\end{equation}
and the leading---${\cal O}(e^2)$---amplitude for Compton scattering
is given by the Feynman diagrams shown in Figure 1 as
\begin{eqnarray}
{\rm Amp}&=&-{e^2/m\over \sqrt{2\omega_i2\omega_f}}\left[\hat{\epsilon}_i\cdot
\hat{\epsilon}_f^*+{1\over m}\sum_n\left({\hat{\epsilon}_f^*\cdot
<0|\vec{p}e^{-i\vec{q}_f\cdot\vec{r}}|n>
\hat{\epsilon}_i\cdot <n|\vec{p}e^{i\vec{q}_i\cdot\vec{r}}|0>\over 
\omega_i+E_0-E_n}\right.\right.
\nonumber\\
&+&\left.\left.{\hat{\epsilon}_i\cdot <0|\vec{p}e^{i\vec{q}_i\cdot\vec{r}}|n>
\hat{\epsilon}_f^*\cdot <n|\vec{p}e^{-i\vec{q}_f\cdot\vec{r}}|0>\over E_0-\omega_f-E_n}
\right)\right]
\end{eqnarray}
where $|0>$ represents the hydrogen ground state having binding energy
$E_0$. 

\begin{figure}
\begin{center}
\epsfig{file=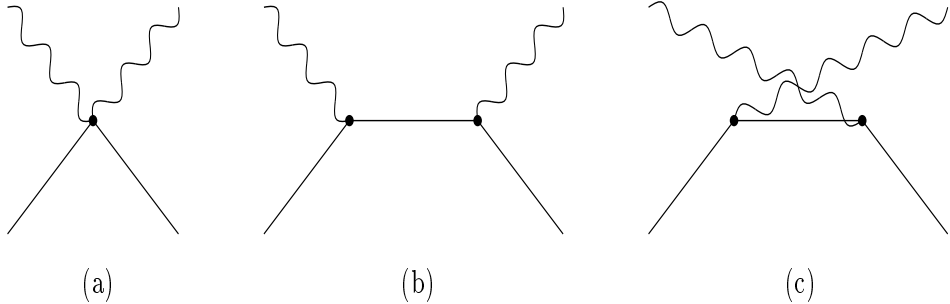,height=4cm,width=12.5cm}
\caption{Feynman diagrams contributing to low energy Compton scattering.}
\end{center}
\end{figure}

\noindent (Note that for simplicity 
we take the proton to be infinitely heavy so
it need not be considered.)  Here the leading component is the familiar 
$\omega$-independent Thomson amplitude and would appear naively 
to lead to an energy-independent
cross-section.  However, this is not the case.  Indeed, as shown in a homework problem,
provided that the energy of the photon is much smaller than a typical 
excitation energy---as is the case for optical photons---the cross section 
can be written as
\begin{eqnarray}
\quad {d\sigma\over d\Omega}&=&
\lambda^2\omega^4|\hat{\epsilon}_{f}^*\cdot
\hat{\epsilon}_{i}|^2\left(1+{\cal O}\left({\omega^2\over (\Delta E)^2}
\right)\right)\label{eq:cc}
\end{eqnarray}
where 
\begin{equation}
\lambda=\alpha_{em}\sum{2|z_{n0}|^2\over E_n-E_0}
\end{equation}
is the atomic electric polarizability,
$\alpha_{em}=e^2/4\pi$ is the fine structure constant, and $\Delta E\sim m
\alpha_{em}^2$ is a typical hydrogen excitation energy.  We note that  
$\alpha_{em}\lambda\sim a_0^2\times {\alpha_{em}\over \Delta E}\sim a_0^3$ 
is of order the atomic volume, as will be exploited below and that the cross 
section itself has the characteristic $\omega^4$ 
dependence which leads to the blueness of the sky---blue light scatters much
more strongly than red.\cite{feyn}

Now while the above derivation is certainly correct, it requires somewhat detailed and
lengthy quantum mechanical manipulations which obscure the relatively simple physics
involved.  One can avoid these problems by the use of effective field theory 
methods outlined above.  The key point is that of scale.  Since the incident photons
have wavelengths $\lambda\sim 5000$A much larger than the $\sim$ 1A atomic size, then at 
leading order the photon is insensitive to the presence of the atom, since the latter
is electrically neutral.  If $\chi$ represents the wavefunction of the atom then the 
effective leading order Hamiltonian is simply that for the hydrogen atom 
\begin{equation}
H_{eff}^{(0)}=\chi^*\left({\vec{p}^2\over 2m}+e\phi\right)\chi
\end{equation}
and there is {\it no} interaction with the field.  In higher orders, 
there {\it can} exist such atom-field interactions and this is
where the effective Hamiltonian comes in to play.  In order to construct the
effective interaction, we demand certain general principles---the
Hamiltonian must satisfy fundamental symmetry requirements.  In
particular $H_{eff}$ must be gauge invariant, must be a scalar under
rotations, and must be even under both parity and time reversal 
transformations.  Also,
since we are dealing with Compton scattering, $H_{eff}$ must be
quadratic in the vector potential.
Actually, from the requirement of gauge invariance it is
clear that the effective interaction should involve only the electric and magnetic
fields
\begin{equation}
\vec{E}=-\vec{\nabla}\phi-{\partial\over \partial t}\vec{A}, 
\qquad \vec{B}=\vec{\nabla}\times\vec{A}\label{eq:ii}
\end{equation}
since these are invariant under a gauge transformation
\begin{equation}
\phi\rightarrow\phi+{\partial\over \partial t}\Lambda,\qquad \vec{A}
\rightarrow\vec{A}-\vec{\nabla}\Lambda
\end{equation}
while the vector and/or scalar potentials are not.  The lowest order
interaction then can involve only the rotational invariants 
$\vec{E}^2,\vec{B}^2$
and $\vec{E}\cdot\vec{B}$.  However, under spatial
inversion---$\vec{r}\rightarrow -\vec{r}$---electric and magnetic
fields behave oppositely---$\vec{E}\rightarrow -\vec{E}$ while
$\vec{B}\rightarrow\vec{B}$---so that parity invariance rules out any
dependence on $\vec{E}\cdot\vec{B}$.  Likewise under time
reversal---$t\rightarrow -t$ we have $\vec{E}\rightarrow \vec{E}$ but
$\vec{B}\rightarrow -\vec{B}$ so such a term is also ruled out by time
reversal invariance.  The simplest such effective Hamiltonian must
then have the form
\begin{equation}
H_{eff}^{(1)}=\chi^*\chi[-{1\over 2}c_E\vec{E}^2
-{1\over 2}c_B\vec{B}^2]\label{eq:ll}
\end{equation}
(Forms involving time or spatial derivatives are much smaller.)
We know from electrodynamics that 
${1\over 2}(\vec{E}^2+\vec{B}^2)$
represents the field energy per unit volume, so by dimensional
arguments, in order to represent an
energy in Eq. \ref{eq:ll}, $c_E,c_B$ must have dimensions of volume.
Also, since the photon has such a
long wavelength, there is no penetration of the atom, so  only classical scattering
is allowed.  The relevant scale must then be atomic size so that we can write
\begin{equation}
c_E=k_Ea_0^3,\qquad c_B=k_Ba_0^3
\end{equation}
where we expect $k_E,k_B\sim {\cal O}(1)$.  Finally, since for photons
with polarization $\hat{\epsilon}$ and four-momentum $q_\mu$ we
identify $\vec{A}(x)=\hat{\epsilon}\exp(-iq\cdot x)$
then from Eq. \ref{eq:ii}, $|\vec{E}|\sim \omega$, 
$|\vec{B}|\sim |\vec{k}|=\omega$ and 
\begin{equation}
{d\sigma\over d\Omega}\propto|<f|H_{eff}|i>|^2\sim\omega^4 a_0^6
\end{equation}
as found in the previous section via detailed calculation.

We see from this example the strength of the effective interaction 
procedure---allowing access to the basic physics with very little formal work. 

\section{Chiral Perturbation Theory}

An important example within the realm of quantum field theory is that of chiral
perturbation theory.  In this case we apply these ideas to the case of QCD.  
Since back in prehistoric times the holy grail of particle/nuclear
physicists has been to construct a theory of elementary particle interactions 
which emulates quantum electrodynamics in that it is elegant, renormalizable, 
and phenomenologically successful.  We now have a theory which satisfies 
two out of the three criteria---quantum chromodynamics (QCD).
Indeed the form of the Lagrangian\footnote{Here the covariant derivative is
\begin{equation}
i D_{\mu}=i\partial_{\mu}-gA_\mu^a {\lambda^a \over 2} \, ,
\end{equation}
where $\lambda^a$ (with $a=1,\ldots,8$) are the SU(3) Gell-Mann matrices,
operating in color space, and the color-field tensor is defined by
\begin{equation}
G_{\mu\nu}=\partial_\mu  A_\nu -  \partial_\nu  A_\mu -
g [A_\mu,A_\nu]  \, ,
\end{equation} }
\begin{equation}
{\cal L}_{\mbox{\tiny QCD}}=\bar{q}(i  {\not\!\! D} - m )q-
{1\over 2} {\rm tr} \; G_{\mu\nu}G^{\mu\nu} \, .
\end{equation}
is elegantly simple, and the theory is renormalizable.  So why are we still not
satisfied?  The difficulty lies with the third criterion---phenomenological 
success.  While at the very largest energies, asymptotic freedom allows the 
use of perturbative
techniques, for those who are interested in making contact with low energy 
experimental findings there exist at least three fundamental difficulties:
\begin{itemize}
\item [i)] QCD is written in terms of the "wrong" degrees of 
freedom---quarks and
gluons---while experiments are performed with hadronic bound states;

\item [ii)] the theory is hopelessly non-linear due to gluon self interaction;

\item[iii)] the theory is one of strong coupling---$g^2/4\pi\sim 1$---so that 
perturbative methods are not practical.
\end{itemize}
Nevertheless, there has been a great deal of recent progress in making 
contact between theory and experiment at low energies using effective 
field theory and chiral symmetry.\cite{eft}

The idea of "chirality" is defined by the projection operators
$\Gamma_{L,R} = {1\over 2}(1\pm\gamma_5)$ which project left- and 
right-handed components of the Dirac wavefunction.
In terms of chirality states the quark component of the QCD Lagrangian
can be written as
\begin{equation} \bar{q}(i\not\! \! D-m)q=\bar{q}_Li\not \! \! D q_L +
\bar{q}_Ri
\not\!\! D q_R -\bar{q}_L m q_R-\bar{q}_R m
q_L \end{equation}
and in the limit in which the light (u,d,s) quark masses are set to
zero
QCD is seen to have an exact $SU(3)_L\times SU(3)_R$ invariance.  Of
course, it is known that the axial part of this symmetry is broken spontaneously
in which case Goldstone's theorem requires the
existence of eight massless pseudoscalar bosons, which couple derivatively
to the rest of the universe.  Of course, in the real
world such massless $0^-$ states do not exist, since 
exact chiral invariance is broken by the small quark mass terms.  
Thus what we have in reality are eight
very light (but not massless) pseudo-Goldstone bosons which make up the
pseudoscalar octet.  Since such states are lighter than their other hadronic
counterparts, we have a situation wherein effective field theory can be 
applied---provided one is working at energy-momenta small compared to 
the $\sim 1$ GeV scale which is typical of hadrons, one can describe the
interactions of the pseudoscalar mesons using an effective Lagrangian.
Actually this has been known since the 1960's, where a good deal of work
was done with a {\it lowest order} effective chiral Lagrangian\cite{gg} 
\begin{equation}
 {\cal L}_2={F_\pi^2 \over 4} \mbox{Tr} (\partial_{\mu}U \partial^{\mu}
 U^{\dagger})+{m^2_{\pi}\over 4} F_\pi^2 \mbox{Tr} 
(U+U^{\dagger})\,  .\label{eq:abc}
\end{equation}
where the subscript 2 indicates that we are working at two-derivative order
or one power of chiral symmetry breaking---{\it i.e.} $m_\pi^2$.
Here $U\equiv\exp(\sum \lambda_i\phi_i/F_\pi)$, where $F_\pi=92.4$ is the pion
decay constant. This Lagrangian is {\it unique}.  It also has predictive 
power.  Expanding to second order in the fields we find the well known
Gell-Mann-Okubo formula for pseudoscalar masses\cite{gmo}
\begin{equation}
3m_{\eta}^2 +m_{\pi}^2 - 4m_K^2 =0 \, \, .
\end{equation}
and is well-satisfied experimentally.  Expanding to fourth order in the 
fields we also reproduce the well-known and experimentally successful
Weinberg $\pi\pi$ scattering lengths.

However, when one attempts to go beyond tree level, in order to unitarize 
the results, divergences arise and that is where the field stopped at the 
end of the 1960's.  The solution, as proposed a decade later 
by Weinberg\cite{wbp} 
and carried out by 
Gasser and Leutwyler\cite{gl}, is to absorb these 
divergences in phenomenological
constants, just as done in QED.  What is different in this case is that
the theory is nonrenormalizabile in that the forms of the divergences are
{\it different} from the terms that one started with.  That means that 
the form of the counterterms that are used to absorb these divergences 
must also be different, and Gasser and Leutwyler wrote down the most general
counterterm Lagrangian that one can have at one loop, involving {\it 
four-derivative} interactions
\begin{eqnarray}
{\cal L}_4 &  =&\sum^{10}_{i=1} L_i {\cal O}_i
= L_1\bigg[{\rm tr}(D_{\mu}UD^{\mu}U^{\dagger})
\bigg]^2+L_2{\rm tr} (D_{\mu}UD_{\nu}U^{\dagger})\cdot
{\rm tr} (D^{\mu}UD^{\nu}U^{\dagger}) \nonumber \\
 &+&L_3{\rm tr} (D_{\mu}U D^{\mu}U^{\dagger}
D_{\nu}U D^{\nu}U^{\dagger})
+L_4 {\rm tr}  (D_{\mu}U D^{\mu}U^{\dagger})
{\rm tr} (\chi{U^{\dagger}}+U{\chi}^{\dagger}
) \nonumber \\
&+&L_5{\rm tr} \left(D_{\mu}U D^{\mu}U^{\dagger}
\left(\chi U^{\dagger}+U \chi^{\dagger}\right)
\right)+L_6\bigg[ {\rm tr} \left(\chi U^{\dagger}+
U \chi^{\dagger}\right)\bigg]^2 \nonumber \\
&+&L_7\bigg[ {\rm tr} \left(\chi^{\dagger}U-
U\chi^{\dagger}\right)\bigg]^2 +L_8 {\rm tr}
\left(\chi U^{\dagger}\chi U^{\dagger}
+U \chi^{\dagger}
U\chi^{\dagger}\right)\nonumber \\
&+&iL_9 {\rm tr} \left(F^L_{\mu\nu}D^{\mu}U D^{\nu}
U^{\dagger}+F^R_{\mu\nu}D^{\mu} U^{\dagger}
D^{\nu} U \right) +L_{10} {\rm tr}\left(F^L_{\mu\nu}
U F^{R\mu\nu}U^{\dagger}\right) \nonumber\\
\end{eqnarray}
where the covariant derivative is defined via
\begin{equation}
D_\mu U=\partial_\mu U+\{A_\mu,U\}+[V_\mu,U]
\end{equation}
the constants $L_i, i=1,2,\ldots 10$ are arbitrary (not determined from chiral
symmetry alone) and
$F^L_{\mu\nu}, F^R_{\mu\nu}$ are external field strength tensors defined via
\begin{eqnarray}
F^{L,R}_{\mu\nu}=\partial_\mu F^{L,R}_\nu-\partial_\nu
F^{L,R}_\mu-i[F^{L,R}_\mu ,F^{L,R}_\nu],\qquad F^{L,R}_\mu =V_\mu\pm A_\mu .
\end{eqnarray}
Now just as in the case of QED the bare parameters $L_i$ which appear
in this Lagrangian are not physical quantities.  Instead the experimentally 
relevant (renormalized)
values of these parameters are obtained by appending to these bare values
the divergent one-loop contributions---
\begin{equation} L^r_i = L_i -{\gamma_i\over 32\pi^2}
\left[{-2\over \epsilon} -\ln (4\pi)+\gamma -1\right]\end{equation}
By comparing predictions with experiment, Gasser and Leutwyler were able 
to determine empirical numbers for each of these ten parameters.  
Typical values are shown in Table 1, together with the way in which they
were determined.
\begin{table}
\begin{center}
\begin{tabular}{l l c}\hline\hline
Coefficient & Value & Origin \\
\hline
$L_1^r$ & $0.65\pm 0.28$ & $\pi\pi$ scattering \\
$L_2^r$ & $1.89\pm 0.26$ & and\\
$L_3^r$ & $-3.06\pm 0.92$ & $K_{\ell 4}$ decay \\
$L_5^r$ & $2.3\pm 0.2$ & $F_K/F_\pi$\\
$L_9^r$ & $7.1\pm 0.3$ & $\pi$ charge radius \\
$L_{10}^r$ & $-5.6\pm 0.3$ & $\pi\rightarrow e\nu\gamma$\\
\hline\hline
\end{tabular}
\caption{Gasser-Leutwyler counterterms and the means by which
they are determined.}
\end{center}
\end{table}
 
The important question to ask at this point is why stop at order 
four derivatives?
Clearly if two-loop amplitudes from ${\cal L}_2$ or one-loop
corrections from ${\cal L}_4$ are calculated, divergences will arise which
are of six-derivative character.  Why not include these?  The answer is that
the chiral procedure represents an expansion in energy-momentum.  Corrections
to the lowest order (tree level) predictions from one 
loop corrections from ${\cal L}_2$
or tree level contributions from ${\cal L}_4$ are ${\cal
O}(E^2/\Lambda_\chi^2)$
where $\Lambda_\chi\sim 4\pi F_\pi\sim 1$ GeV is the chiral scale\cite{sca}.
Thus chiral
perturbation theory is a {\it low energy} procedure.  It is only to the extent
that the energy is small compared to the chiral scale that it makes sense to
truncate the expansion at the one-loop (four-derivative) level.  Realistically this
means that we deal with processes involving $E<500$ MeV, and, 
for such reactions the procedure is found to work very well.

In fact Gasser and Leutwyler, besides giving the form of the ${\cal O}(p^4)$ chiral
Lagrangian, have also performed the one loop integration and have written the
result in a simple algebraic form.  Users merely need to look up the result in
their paper and, despite having ten phenomenological constants, the theory is
quite predictive.  An example is shown in Table 2, where predictions are
given involving quantities which arise using just two of the 
constants---$L_9,L_{10}$.  The table also reveals at least one intruguing
problem---a solid chiral prediction, that for the charged pion 
polarizability, is possibly violated although this is not clear since there 
are three experimental results, only one of which is in disagreement.  
Clearing up this discrepancy should be a focus of future experimental work.  
Because of space limitations we shall have to be
content to stop here, but interested readers can find applications to other
systems in a number of review articles\cite{cptr}.

\begin{table}
\begin{center}
\begin{tabular}{cccc}\hline\hline
Reaction&Quantity&Theory&Experiment\\
\hline
$\pi^+\rightarrow e^+\nu_e\gamma$ & $h_V(m_\pi^{-1})$ & 0.027 
& $0.029\pm 0.017$\cite{pdg}\\
$\pi^+\rightarrow e^+\nu_ee^+e^-$ & $r_V/h_V$ & 2.6 & $2.3\pm 0.6$\cite{pdg}\\
$\gamma\pi^+\rightarrow\gamma\pi^+$ & $(\alpha_E+\beta_M)\,(10^{-4}\,{\rm fm}^3)$& 0
&$1.4\pm 3.1$\cite{anti}\\
      &$\alpha_E\,(10^{-4}\,{\rm fm}^3)$&2.8 & $6.8\pm 1.4$\cite{anti1}\\
 & & & $12\pm 20$\cite{russ}\\
 & & & $2.1\pm 1.1$\cite{slac}\\
\hline
\end{tabular}
\caption{Chiral Predictions and data in radiative pion processes.}
\end{center}
\end{table}
 
\section{NN Effective Field Theory}

Prompted in part by the success of chiral effective field theories, a
number of groups have attempted to extend such programs to the arena
of NN scattering.  In the standard approach, of course, one assumes
the validity of some potential form, chooses parameters in order to
match scattering experiments in some energy region and then attemptes 
to predict phase shifts at other energies.  The idea in EFT is
similar---at low energies one should be able to express scattering
only in terms of observables such as the scattering length and
effective range.  In terms of potentials, this means that one is 
characterizing the scattering in terms of the quantities
\begin{equation}
a\simeq -{m_r\over 2\pi}U(\vec{p}=0),\quad r_e\simeq {1\over
3a^2}{m_r\over 2\pi}\vec{\nabla}^2U(\vec{p}=0)
\end{equation}  
{\it i.e.} in terms of the low energy momentum space potential
$U(\vec{p})=\int d^3r\exp(i\vec{p}\cdot\vec{r})V(r)$.  This
suggests the use of simple contact interactions which reproduce these
results, and that is the approach used by EFT practitioners.  If one
is interested only in NN scattering the effective range and EFT
results are identical.  However, if one wishes to include interactions
with external fields, the simple EFT Lagrangian must be supplemented
by various undetermined counterterms\cite{crs}.  For example, in the case of
coupling to an electromagnetic field at lowest order one must include
two magnetic counterterms, usually called $L_1,L_2$\cite{ssw}.  The coupling
$L_2$ is found from comparison of the experimental deuteron magnetic
moment---0.857 nm---with its simple one-body
value---$\mu_p+\mu_n=0.88$ nm.  The coupling $L_1$ is determined by
comparison of the measured threshold cross section for the radiative
capture process $n+d\rightarrow d+\gamma$---$\sigma_{exp}=
334.5\pm 0.5$ mb for and
incident thermal neutron velocity 2200 m/s---with the lowest order
(zero range approximation) theoretical one body operator value
\begin{equation}
\sigma_{LO}={2\pi\alpha\over v}(\mu_p-\mu_n)^2\left({\gamma\over
M_N}\right)^5a_s^2(1-{1\over \gamma a_s})^2=297.2\,\,{\rm mb}
\end{equation}
where $\alpha$ is the fine structure constant, $a_s=-23.7$ fm is the
${}^1S_0$ scattering length, and $\gamma=\sqrt{M_NB}=45.6$ MeV, with
$B$ being the deuteron binding energy.  The ten percent discrepancy
between the experimental and one body values for the cross section is
well known and can be understood in a conventional nuclear physics
approach in terms of a combination of pion exchange and $\Delta$
contributions\cite{ris}.  In the EFT scheme, the origin of the discrepancy is
irrelevant.  One merely fixes the counterterm $L_1$ in order to
reproduce the experimental cross section.  At this level then it
appears that there is no predictive power.  However, this is illusory
since once the cross section is determined at threshold the energy
dependence is predicted unambiguously and agreement is excellent as
shown in Figure 2.  Higher order contributions have also been
calculated and convergence is good at low energies, resulting in a 1\%
prediction for the low energy cross section, which is a critical
ingredient into nucleosynthesis calculations\cite{rup}.  Of course, having
determined the value of the counterterm $L_1$ in this way, it can now
be used i) in order to compare with theoretical values, ii) to predict
the inverse cross section for deuteron photodisintegration
$\gamma+d\rightarrow n+p$, iii) or for deuteron electrodisintegration
$e+d\rightarrow e+n+p$, etc.

\begin{figure}
\begin{center}
\epsfig{file=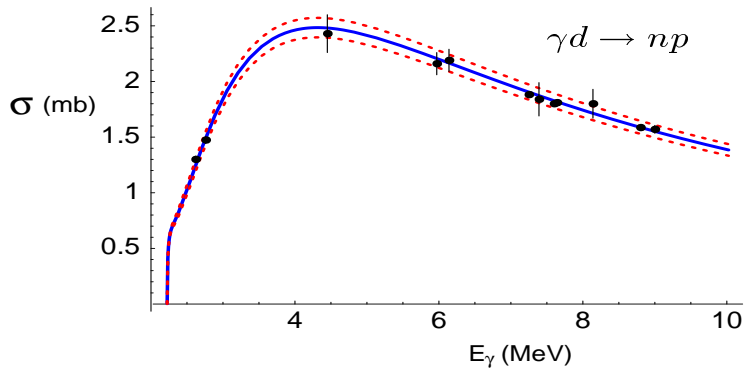,height=5cm,width=10cm}
\caption{The photodissociation cross section for $\gamma+d\rightarrow
n+p$ as a function of energy with counterterm $L_1$ determined via the
threshold capture cross section for $n+p\rightarrow d+\gamma$.  The
dashed lines indicate the theoretical uncertainty.}
\end{center}
\end{figure}   

\section{Calibrating the Sun}

Another application of EFT methods is in weak interactions.  
The mechanism by which stars---especially our sun---generate their 
prodigious energy has long been of interest to physicists.  First
proposed at the end of the 1930's by Bethe and Critchfield, 
the basic process is summarized via
\begin{equation}
4p\rightarrow {}^4He+2e^++2\nu_e+25\,\,{\rm MeV}
\end{equation}
while the detailed reaction picture is shown in Figure 1.  However,
despite the fact that this hypothesis has been around now for over
six decades, it is only recently, with the advent of large scale
neutrino detectors, that direct tests of this picture have become
possible.  As is well-known, such tests have in general revealed a
deficit of such neutrinos and this problem has given rise to the
suggestion of neutrino oscillations, which has become a field unto
itself\cite{hh}.  Implicit in this observation is the assumption that the
rate for the reaction 
\begin{equation}
p+p\rightarrow d+e^++\nu_e\label{eq:so}
\end{equation}   
which begins this chain is correctly calculated.  A theoretical
evaluation is required here since the only experimental measurement---
\begin{equation}
\sigma(\nu_e+d\rightarrow p+p+e^-)_{exp}=
(53\pm 18)\times 10^{-42}\,\,{\rm cm}^2
\end{equation}
by LAMPF E31\cite{la}---while in agreement with the corresponding theoretically
calculated value\cite{kn}
\begin{equation}
\sigma(\nu_e+d\rightarrow p+p+e^-)_{th}=52\times 10^{-42}\,\,{\rm cm}^2,
\end{equation}
has only $\sim 35\%$ precision.

On the theoretical side, calculation of the reaction Eq. \ref{eq:so}
consists of two components:
\begin{itemize}
\item [i)] convolution of the one-body operator, obtained from neutron
beta decay, with the deuteron and pp wavefunctions;
\item [ii)] evaluation of the two-body piece, which is generally
described via a meson-exchange approach.
\end{itemize}
The first part is reasonably secure, which cannot be said for
the two-body part of the calculation.  This is evidenced by the 
fact that two such calculations of total cross sections for
the related reaction
\begin{equation}
\nu_x+d\rightarrow n+p+\nu_x\label{eq:sno}
\end{equation}
by Ying, Haxton, and Henley (YHH)\cite{yhh} and by Kubodera and Nozawa
2(KN) \cite{kn} differ by $\sim 5\%$, as shown in Figure 3.
YHH included exchange currents only to the extent that they 
are incorporate through the constraints of current conservation,
imposed by a generalized Siegert's theorem.  KN included in
addition model-dependent contributions of exchange currents.

\begin{figure}
\begin{center}
\epsfig{file=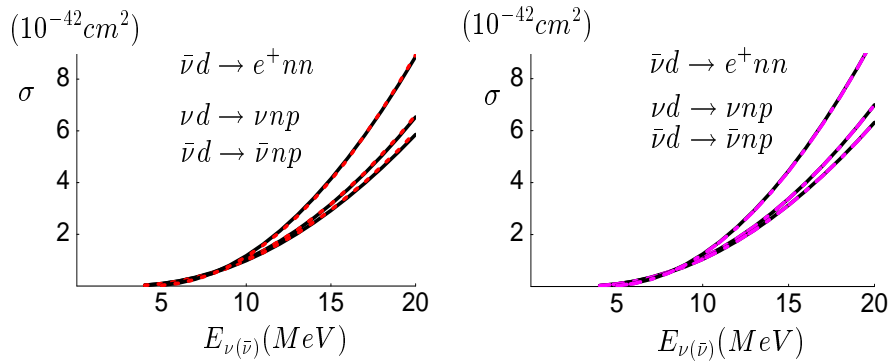,height=5cm,width=12cm}
\caption{Inelastic $\nu(\bar{\nu})d$ cross sections vs. incident
$\nu(\bar{\nu})$ energy.  The solid curves in the left hand figure are
the results of Kubodera and Nozawa, while the dot-dashed curves, which
sit on top of the solid curves are NLO in EFT with $L_{1,A}$=6.3 fm$^3$.
The results given in the right hand graph are those of Ying, Haxton,
and Henley, with the dashed curves which lie on top generated at NLO
in EFT with $L_{1,A}$=1.0 fm$^3$.}
\end{center}
\end{figure}

    There has been recent work in which model-dependence of the
exchange currents, arising primarily from poorly known
couplings involving N-$\delta$ transitions, has been reduced
by appealing to the known $\beta$ decay rate of the similar
Gamow-Teller transition in $^3$He.  This, of course, does not
entirely circumvent the problem because three-body nuclear
forces are then entangled in the analysis, and because a
specific model-dependent set of exchange current diagrams 
are considered. 
     The YHH and KN calculations are both fairly traditional
nuclear physics calculations, and the resulting differences 
are probably a fair estimate of the uncertainties arising from
the choice of nuclear potential and the description of 
exchange current and other short-range effects.  While such 
discrepancies are perhaps not important
for some applications, in the case of the reaction Eq. \ref{eq:so},
which initiates the solar burning chain, and Eq. \ref{eq:sno}, which
is used as a signal for neutrino-initiated neutral current reactions 
at the SNO detector, such a considerable uncertainty is unacceptable.  

The methods of effective field theory can be used to resolve the problem.  
The inputs required for such calculations 
are simply the neutron decay
axial decay constant $g_A$ and the experimental value of the 
deuteron binding energy, together with the experimental $np,\,pp$ 
scattering lengths and effective ranges.  Since 
this is a low energy reaction the rescattering corrections which
renormalize the basic decay process can be reliably included, and
such calculations for the charged and neutral current neutrino
reactions have recently been done by Butler and Chen\cite{bc}.  A
parallel and related calculation for the reaction Eq. \ref{eq:so} has
been performed by Kong and Ravndal\cite{kr}.     

In both evaluations, however, there exists the same unknown
quantity---a four-nucleon axial current counterterm of the form
\begin{equation}
\vec{A}^{(2)}={1\over 8}L_{1A}(N^T\sigma_2\vec{\sigma}\tau_2N)^\dagger
(N^T\sigma_2\tau_2\tau_-N)
\end{equation}
where $L_{1A}$ has units of fm$^3$ and is expected by is by EFT
scaling arguments to be in the range $-6\,\,{\rm fm}^3\leq L_{1A}
\leq +6\,\,{\rm fm}^3$,
corresponding to a cross section difference of $\sim 10\%$.  In fact,
Butler and Chen have shown that the calculations of YNN and KN can
be completely characterized via the values  
\begin{equation}
L_{1A}=\left\{\begin{array}{cc}
6.3 \,\,{\rm fm}^3& YHH\cite{yhh}\\
1.0 \,\,{\rm fm}^3& KN\cite{kn}
\end{array}\right.
\end{equation}
This large difference for values of $L_{1A}$ simply corresponds to differing
assumptions about the short-range properties of the nucleon-nucleon
interaction associated with the meson-exchange calculations.

Obviously, one can generate alternative values of $L_{1A}$ via
other descriptions of short-range structure, but it is clear that the best
approach would be to measure this counterterm experimentally, say via  
the charged current reaction
\begin{equation}
\nu_e+d\rightarrow p+p+e^-
\end{equation}
The corresponding experimental value of the counterterm could then be used to
yield a theoretically rigorous $\sim 2\%$ prediction for the reaction
rate for the pp neutrino process which initiates the solar burning
cycle {\it and} for the neutral current deuteron disintegration
reaction Eq. \ref{eq:sno} which is used in the SNO detector in order
to indicate that a neutral current neutrino reaction has occurred.
Finally, it is clear that careful measurement to pin down the size of the
counterterm $L_{1A}$  would offer a target for theorists
to shoot at in order to gauge their models of short distance NN
structure. 

\section{Conclusions}

We have tried to give a didactic introduction to effective field
theory methods in a variety of applications.  In each case we have
generated an effective interaction which is valid in a given (usually
low energy) range.  In the case of Rayleight scattering the EFT
procedure was found to reveal the physics of the process in a simple
fashion.  In the case of low energy QCD, chiral perturbation theory was
able to elicit reliable predictions in an otherwise intractable energy
region.  In the case of the NN interaction we have seen how the use of
EFT methods allows model-independent predictions for low energy weak
and electromagnetic processes to be made once low energy counterterms
are determined theoretically.  Space limitations do not allow us to
give further examples, but work is underway in a number of areas.  One
is the extension to systems with three or more constituents.  In this
case, Bedaque, Hammer, and van Kolck have shown how the $nd$
scattering length in the quartet channel $a_{3\over 2}^{exp}=6.35\pm
0.02$ fm can be predicted in terms of known NN quantities, yielding
$a_{3\over 2}^{eft}=6.32\pm 0.1$ fm\cite{bhv}.  However, things are more
difficult in the doublet case.  Efforts are also underway to simplify
large basis shell model calculations using EFT methods.  Another
challenge is how to extend such techniques to higher energy.  Here the
counting scheme which should work, PDS wherein pion exchange is
treated perturbatively, seems to have convergence difficulties\cite{pds}, while 
that which is less on firm ground---Weinberg power counting---seems to  
give reasonable results\cite{wb}.  There is still much to do!
 
\begin{center}
{}\bf Acknowledgement
\end{center}

It is a pleasure to acknowledge the hospitality of the organizers of
this Few Body meeting.  This work was supported in part by the
National Science Foundation.

\end{document}